\documentclass{v23tmex}

\def\be{\begin{equation}}
\def\ee{\end{equation}}
\def\bea{\begin{eqnarray}}
\def\eea{\end{eqnarray}}

\newcommand{\Photo}{}

\begin{document}
\vspace*{4cm}
\title{THE PEVATRONS AND THE HAWC OBSERVATORY IN MEXICO \footnote{Accepted to be published in Theory meeting experiments (TMEX-2023), Vietnam, 5-11 January 2023.}}

\author{E. DE LA FUENTE \footnote{Also Doctorado en Ciencias F\'isico--Matem\'aticas, CUValles, Universidad de Guadalajara, Carretera Guadalajara--Ameca Km. 45.5, 46600, Ameca, Jalisco.}}

\address{Departameto de F\'isica, CUCEI, Universidad de Guadalajara\\
Blvd. Marcelino Garc\'ia Barrag\'an 1421 esq. Calzada Ol\'impica, Guadalajara, Jalisco, M\'exico }

\author{ J.L. FLORES, G. GARC\'IA-TORALES}

\address{Departamento de Fot\'onica, CUCEI, Universidad de Guadalajara\\
Blvd. Marcelino Garc\'ia Barrag\'an 1421 esq. Calzada Ol\'impica, Guadalajara, Jalisco, M\'exico }

\author{ J.C. D\'IAZ-V\'ELEZ}

\address{Dept. of Physics and Wisconsin IceCube Particle Astrophysics Center, University of Wisconsin–Madison, Madison, WI, 53706, USA}

\author{THE HAWC COLLABORATION \footnote{See at the end of this proceeding and authorlist at \url{https://www.hawc-observatory.org/collaboration/}.}}
\address{and}

\author{ I. TOLEDANO-JU\'AREZ }

\address{Doctorado en Ciencias F\'isicas, CUCEI, Universidad de Guadalajara \\
Blvd. Marcelino Garc\'ia Barrag\'an 1421 esq. Calzada Ol\'impica, Guadalajara, Jalisco, M\'exico }

\maketitle\abstracts{
The discovery of ultra-high energy gamma-ray sources (detected at energies $\geq$ 100 TeV) thanks to highly sensitive observatories such as the High Altitude Water Cherenkov (HAWC) Observatory, the Tibet AS-gamma Experiment, and the Large High Altitude Air Shower Observatory (LHAASO), marked the beginning of the sub--PeV and PeV era in gamma-ray astrophysics (energies $\sim$ 0.1 to 100 PeV). This new astrophysics is closely related to the emerging topic of PeVatrons, in which HAWC has played a remarkable role and will continue to contribute discoveries and relevant studies thanks to the better data provided by the installed outriggers. In this paper, we present a brief overview of the PeVatrons and the HAWC observatory. }



\section{Introduction}

It is known that nature can accelerate particles to energies up to hundreds of exa-electronvolts or EeVs (DeAngelis \& Pimienta \cite{DeAngelis2018}), while the most powerful man-made particle accelerators can reach up to 14 TeV (in a collision; e.g., the Large Hadron Collider \footnote{\url{https://fcc-cdr.web.cern.ch/}}). The best place to study the Universe at higher energies is nature itself (e.g., violent and cataclysmic events associated with black holes, binary systems, and the death of massive stars: the extreme and violent Universe; e.g. Sinnis \cite{Sinnis2021}). gamma--ray astrophysics offers the best opportunity to study this particle acceleration, especially with observations ranging from 100 MeV to over 100 TeV. Observations can be made with ground-based observatories, using Vavilov-Cherenkov radiation in air or water (e.g. Kobzev \cite{Kobzev2010}), plastic scintillators (e.g. Leo \cite{Leo1994}), and/or a combination of these (e.g., Ma et al. \cite{Ma2022}). 
Despite the fact that Earth's atmosphere absorbs these high-energy particles, it is possible to build an \textit{ad hoc} instrument and use the atmosphere as a calorimeter to detect the secondary particles produced by this absorption (the extensive air showers or EAS phenomena).


Astroparticles include gamma rays (photons with the highest energies), cosmic rays (mostly relativistic protons that fill the galactic disc and are part of the interstellar medium), and neutrinos (electrically neutral fermions with very low mass). The field of multi--messenger astronomy combines the study of astroparticles, gravitational waves, and multi--wavelength astronomy. When gamma rays and primary cosmic rays from the cosmos (protons and heavier atomic nuclei) enter the Earth's atmosphere (at an altitude of $\sim$ 30 km), an EAS is produced. There are two (simultaneous) types of EAS: one due to cosmic rays with three components (strong, weak or electromagnetic, and hadronic) and one due to gamma rays (electromagnetic component only). These components produce a cascade of secondary particles that propagates through the atmosphere until they reach the ground. In the case of cosmic rays, the strong component produces positive and neutral kaons, that decay into charged and neutral pions. Charged pions decay into muons (which produce positrons, electrons, and muonic atmospheric neutrinos), and neutral pions decay into two lower--energy gamma rays generating a positron and an electron by pair production. The electromagnetic component is the shower of secondary particles associated with the electrons and positrons, and the hadronic component produces nuclear fragments such as protons, neutrons, charged pions, and kaons. In gamma rays, the primary particle interacts with a nucleus in the air, and produces a secondary positron and electron through pair-production (no neutral pion decay involved). Positrons and electrons interact with atmospheric nuclei and produce further gamma rays with lower energy by Bremsstrahlung radiation, which decay into further positrons and electrons. This cyclic process continues until positrons and electrons reach the ground. To complete this description with details, see section 1 of de la Fuente et al.\cite{delaFuente2013}

The secondary particles are those detected at the ground by plastic scintillator detectors (e.g., Tibet AS--gamma Observatory; see this conference volume; Amenomori et al.\cite{Amenomori2019} and references therein) or which produce the Vavilov--Cherenkov radiation \footnote{When a particle with a speed greater than the speed of the light (V$_L$) in a medium like air (V$_L$$\sim$0.99c) or water (V$_L$=0.75c), crosses this medium, a optical radiation in the UV range (peak about 4200\AA) is produced.} detected in the air (Imaging Air Cherenkov Telescopes or IACTs such as the Cherenkove Telescope Array or CTA \footnote{See this conference volume and \url{https://www.cta-observatory.org/}}, The High Energy Stereoscopic System or H.E.S.S. \footnote{see references at \url{https://www.eoportal.org/other-space-activities/hess\#hess-telescopes}}, Major Atmospheric gamma-ray Imaging Cherenkov Telescope or MAGIC \cite{Lorenz2005}, and the Very Energetic Radiation Imaging Telescope Array System or Veritas \footnote{see \url{https://veritas.sao.arizona.edu/}}) or in water (Water Cherenkov Detectors or WCDs such as the High Altitude Water Cherenkov Observatory or HAWC \cite{Abeysekara2023}), or hybrids like the Large High Altitude Air Shower Observatory or LHAASO \cite{Ma2022}. The Vavilov--Cherenkov light is quantified using sensors such as photomultipliers tubes or PMTs installed in the detectors. Theory about EAS can be found in Engel et al.\cite{Engel2011}. Therefore, in observing the highest energies, the Earth's atmosphere can be considered the upper detector that produces the EAS, and an instrument that detects and quantifies the secondary particles is the lower detector. The key to conducting cosmic ray and gamma--ray studies is to distinguish between the particle showers produced by cosmic ray interactions (which are rich in muons) from those produced by gamma--ray interactions (consisting mostly electron--positron pairs). Each observatory and telescope \footnote{Telescope means that the instrument can be pointed directly at the sky, and an observatory means that the instrument is fixed and crosses the sky} has its own discrimination technique, and by studying the EAS using simulations, it is possible to determine the energy of the primary signal. For example, in general way, if the original energy in the natural particle accelerator (equivalent to the energy beam of a human-manufactured particle accelerator) was up to 1 PeV (cosmic ray), the energy of the detected primary gamma ray is up to 100 TeV.

The recently inaugurated LHAASO observatory (see this conference volume) is the actual observatory that can observe PeV energies. In contrast, observatories such as HAWC and Tibet AS-gamma are sub--PeV observatories. Upgrades of LHAASO (see Ma et al.\cite{Ma2022} and this conference volume), detectors such as CTA in the northern and southern hemispheres, and southern observatories such as the The Southern Wide-field gamma-ray Observatory or SWGO (e.g., Hinton et al.\cite{Hinton2021}) and the Andes Large area PArticle detector for Cosmic ray physics and Astronomy or ALPACA (e.g., Okukawa et al.\cite{Okukawa2023}, Kato et al.\cite{Kato2021}, and references therein), will help reveal the Universe in the sub--PeV to PeV range with greater accuracy and provide answers to hot open topics such as the PeVatrons; natural PeV particle accelerators, which are considered unique laboratories that can help answer open questions about cosmic rays such as: What are the sources, how can we identify them, what is the acceleration mechanism, what are the effects on propagation, anisotropy in the arrival directions, what is the chemical composition, and what is the maximum enerrgy the nature can accelerate particles assuming that the knee structure at $\sim$ 4 PeV corresponds to the maximum energy of cosmic rays accelerated in the Galaxy.


\section{The HAWC observatory and the PeVatrons}

\begin{figure}
\centerline{\includegraphics[width=0.92\linewidth]{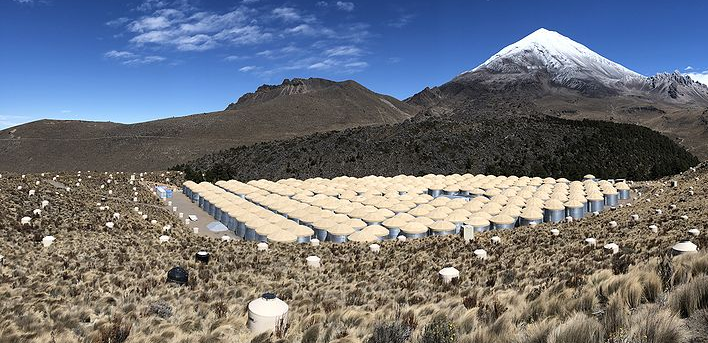}}
\centerline{\includegraphics[width=0.92\linewidth]{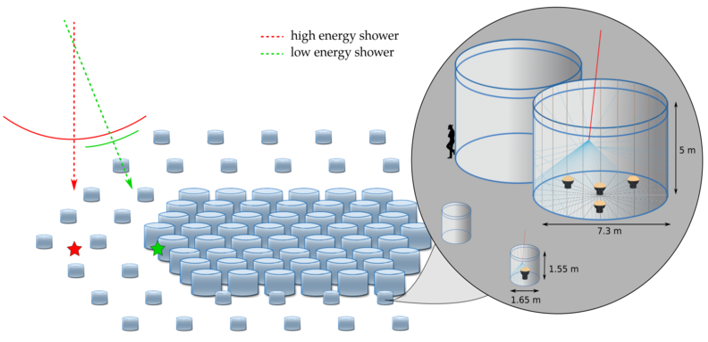}}
\caption[]{Top. The HAWC site with the primary detector (near the centre) surrounded by the outriggers. Bottom. Schematic diagrams of the WCDs for the primary detector (large WCDs) and the outriggers (small WCDs). See Abeysekara\cite{Abeysekara2023} and Marandon\cite{Marandon2021} for details.}
\label{fig:Figure1}
\end{figure}

\subsection{The HAWC observatory}

HAWC, the successor of the Milagro Observatory, is a second-generation high-sensitivity WCD located on the slope of Sierra Negra volcano in Puebla, Mexico (18.99$^{\circ}$ North and 97.31$^{\circ}$ West) at an altitude of 4100 m.a.s.l. It consists of a primary detector surrounded by a set of outriggers. The primary detector encloses 300 WCDs covering $\sim$ 22000 m$^2$. Each WCD is a steel tank 7.3 m in diameter and 4.5 m high, enclosing a bladder containing $\sim$ 180,000 liters of purified water. At the bottom of each WCD, a ten-inch Hamamatsu R7081 is surrounded by three eight-inch Hamamatsu R5912 PMTs in an equilateral triangle configuration. The secondary particles from the EAS pass through the water of each WCD and produce a Vavilov-Cherenkov light that the PMTs detect. The information is collected via a cable network and forwarded to the electronics room, where the information is stored and processed. 

\begin{figure}
\centerline{\includegraphics[width=\linewidth]{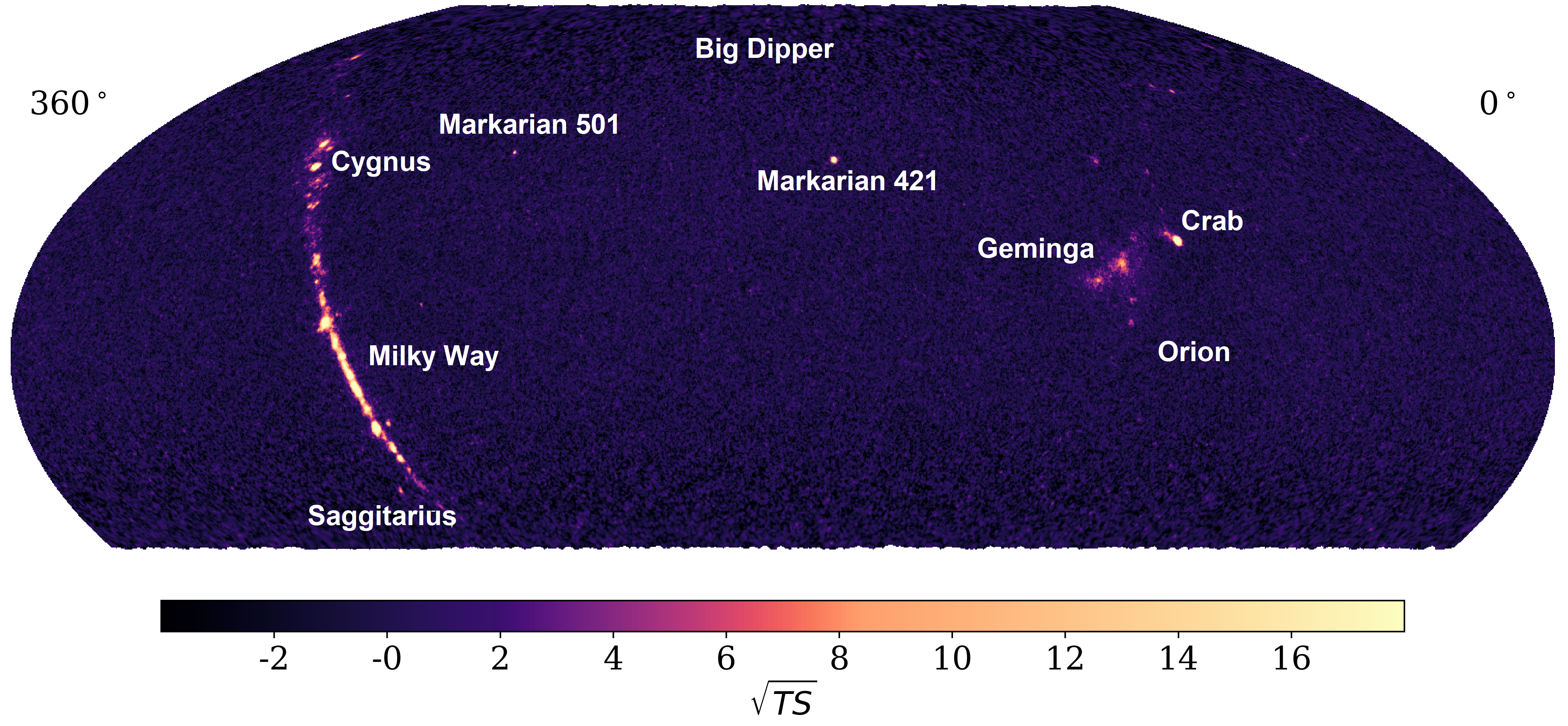}}
\centering
\caption[]{HAWC equatorial coordinate map from 100 GeV to beyond 100 TeV. The data cover the period from $\sim$ Jun. 2015 to Jun. 2022, corresponding a total lifetime of about 2400 days. Constellations and object of interest are marked.}
\label{fig:Figure2}
\end{figure}

\begin{figure}
\includegraphics[width=\linewidth]{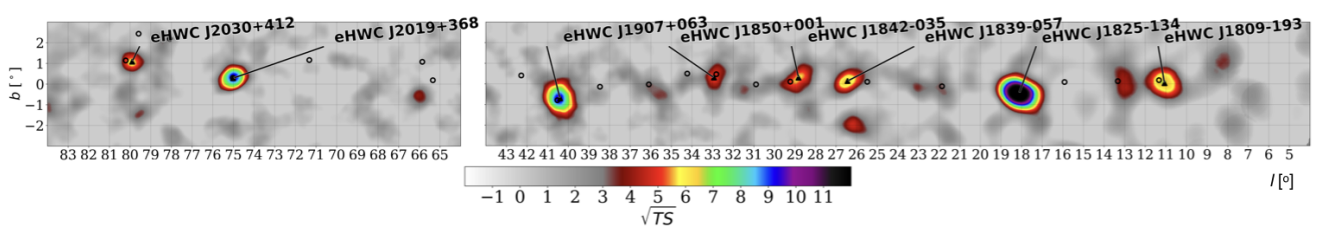}
\caption[]{HAWC maps for several galactic sources with emission above 56 TeV. The data cover the period from $\sim$ Jun. 2015 to Jul. 2018, corresponding a total lifetime of about 1039 days.  See Abeysekara\cite{Abeysekara2020} for details.}
\label{fig:Figure3}
\end{figure}

\begin{figure}
\includegraphics[width=\linewidth]{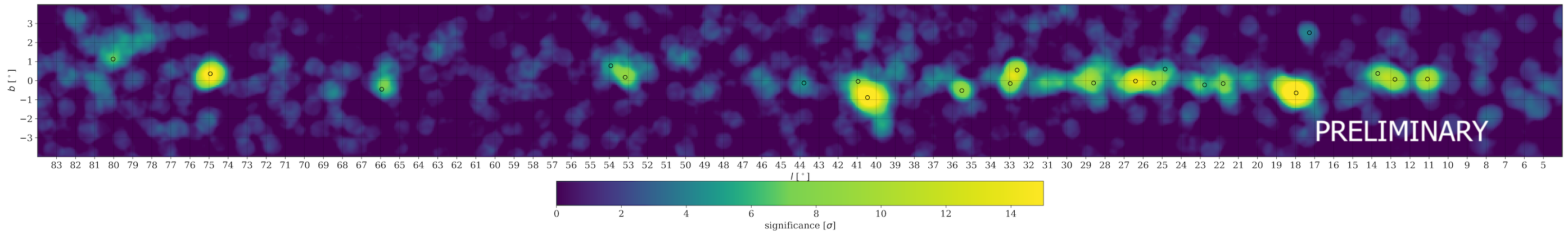}
\includegraphics[width=\linewidth]{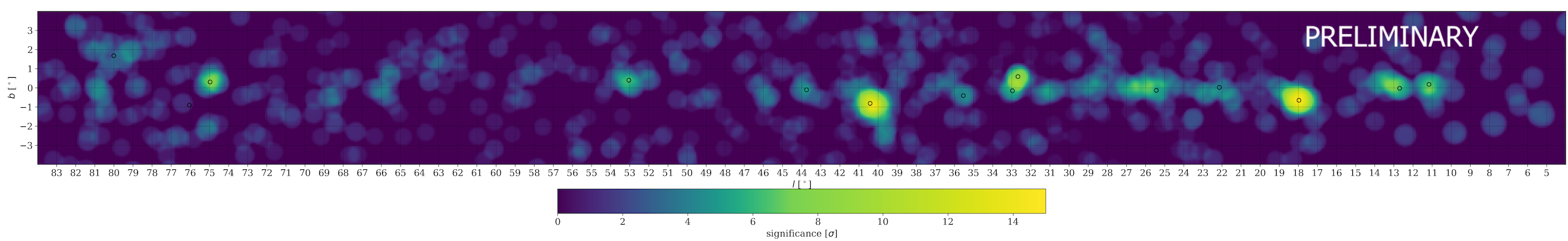}
\includegraphics[width=\linewidth]{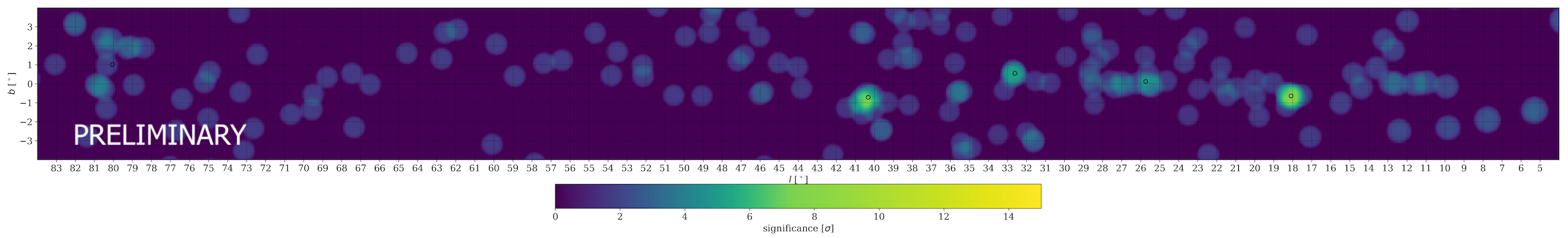}

\caption[]{Preliminary HAWC 0.5 degree extended maps of the galactic centre with sources above 56 TeV (top) 100 TeV (middle) and above 177 TeV (bottom). The livetime of the maps is 2400 days. The brightest source near to the center of each map is the source 3HWC J1908+063.}
\label{fig:Figure4}
\end{figure}

After testing with the VAMOS engineering array (2010--2011), it took $\sim$ 3.5 years to complete the primary detector, which was inaugurated on 20 March 2015. In Figure~\ref{fig:Figure1}, we show the HAWC observatory with the primary detector and outriggers array, and Figure~\ref{fig:Figure2} shows the sky map from 100 GeV to beyond 100 TeV. The most up-to-date catalog paper is the third catalog (Albert et al.\cite{Albert2020}), and in context, it was a pioneering observatory in the discovery of TeV--halo objects (e. g., Abeysekara et al. \cite{Abeysekara2017}). Furthermore, HAWC proved to be an excellent observatory for sub--PeV studies with the discovery of sources above 56 TeV in 2020 (see Figure~\ref{fig:Figure3} and Abeysekara \cite{Abeysekara2020}). Additional data has led to preliminary discoveries of even more sources above 100 TeV and 177 TeV (Figure.~\ref{fig:Figure4}). The technical details of the primary detector, the history, the science summary, and the future of the HAWC observatory are presented in Abeysekara et al. \cite{Abeysekara2023}. Except for Figure ~\ref{fig:Figure3}, the livetime of the data presented in this work is $\sim$ 2400 days.

A new HAWC era began in 2023 with the outriggers array of 345 WCDs installed between 2016 and 2018, improving the sensitivity of the primary detector above 10 TeV and its angular resolution by a factor of 3 to 4. Each outrigger's WCD is a commercial plastic tank with a diameter of 1.55 m and a height of $\sim$ 1.65 m, filled with $\sim$ 1100 liters of purified water and equipped with Hamamatsu R5912 eight-inch PMTs upward facing and anchored to the ground. The distance between the WCDs is between 12 and 18 m, extending the range of the primary detector to $\sim$ 40,000 m$^2$ from the centre. The details and more information about the outriggers can be found at Marandon et al.\cite{Marandon2021}

\subsection{PeVatrons}

The term \textit{PeVatron} was coined in 2016 by the H.E.S.S. Collaboration through a study of the Galactic centre \cite{Abramowski2016}. Although MAGIC and Veritas performed previous IACTs observations in 2011 and 2016, the work of the H.E.S.S. Collaboration can be considered the first robust prediction of a very high energy (0.1 to 100 TeV) cosmic hadron accelerator in the Galaxy. In this work, the term PeVatron, associated with the search for the origin of PeV cosmic rays, refers to an accelerator for cosmic rays beyond 1 PeV. The era of PeVatrons formally began in 2021 with the sub--PeV to PeV observations of Tibet AS-gamma, HAWC (March 2021), and LHAASO experiments culminating with the first LHAASO catalog of 12 PeVatrons candidates (May 2021), including Cygnus OB2, the supernova remnant (SNR) G106.3+2.7, and the Crab (Cao et al.\cite{Cao2021}). 

HAWC has shown that massive star clusters such as Cygnus OB2 associated with the Cygnus cocoon (Ackerman et al.\cite{Ackermann2011}) can be considered PeVatrons (Abeysekara \cite{Abeysekara2021}), not just SNRs, the Galactic centre, pulsar wind nebulae (PWN) and associated TeV--halo objects (leptonic emitters). This result has profound implications as it also establishes a (possible) link between the formation of massive stars and very (or ultra?) high gamma-ray emission. The Tibet AS-$\gamma$ experiment studied the SNR G106.3+2.7 (Amenomori \cite{Amenomori2021}), suggesting that this object is a hadronic and enigmatic SNR, in agreement with theoretical models like Fang et al.\cite{Fang2022}. This result opened an intense debate on hadronic PeVatrons associated with supernova remnants and cosmic rays. In 2021, LHAASO identified the Crab Nebula as a leptonic PeV emitter (Aharonian et al.\cite{Aharonian2021}), suggesting that a reference in gamma--ray astronomy and perhaps the first PeVatron reported by LHAASO is leptonic and not hadronic. This result is relevant and vital because of the considerations: Is the Crab is a PeVatron (?), and if so, what is its role in the hadronic context (?). Other fundamental open questions are 1.- What is the definition of a PeVatron, 2.- what is the origin and nature of PeVatron emission: hadronic (protons) or leptonic (electrons), and 3.- what is the multi-wavelength counterpart of a PeVatron. Some answers may be forthcoming at the International Conference on Cosmic Rays (ICRC) in 2023. In the meantime, we point to the actual conflict: determining the nature (physical mechanism) of a PeVatron: hadronic (with protons and gamma rays produced by the decay of neutral pions created in the interaction of protons with molecular clouds), leptonic (with electrons and gamma rays produced for example by the inverse Compton effect) or co-acceleration (both hadronic and leptonic).  

Although the association of a gamma--ray source with a PeVatron is probed with the gamma--ray spectrum in TeV emissions, where a cut-off of a few hundred TeV is expected for cosmic rays accelerated to PeV energies (e.g., Gabici \& Aharonian \cite{Gabici2007}), in the PeV range, it is impossible to distinguish between leptonic and hadronic models in the spectral energy distribution (SED). Since the beginning of the study of PeVatrons, hadrons have been preferred. There is a historical and general reason for this: the context of cosmic rays (e.g., Celli et al. \cite{Celli2020} and references therein).  Theoretical models consider (young; few hundred years) supernova remnants as PeVtrons; hadrons accelerated with energies up to the knee of the cosmic ray spectrum ($\sim$ 3 PeV; e.g., Baade \& Zwicky \cite{BZ1934}). This fact is confirmed by the evidence of gamma--ray emissions produced by the decay of neutral pions (e.g., Ackermann et al.\cite{Ackermann2013} and references therein). Therefore, the favored is hadronic, but the observed cannot discard the leptonic. Moreover, the evidence for PeV cosmic-ray acceleration has not been well-established in the observations of individual gamma--ray sources from the first LHAASO catalog. The second LHAASO catalog could show whether it can distinguish between leptonic and hadronic emissions in the higher PeV energy range. Nevertheless, one way to distinguish between the two emissions is to combine the sub--PeV and PeV emissions with X--ray and Fermi--LAT observations at SED.

Furthermore, since the 1960s,  X-ray astronomy has discovered SNRs and PWNe as common objects dominated by leptonic emission (e.g., inverse Compton), and SNR RX J1713.7-3946 shows hadronic and leptonic components (e.g., Cristofari et al.\cite{Cristofari2021} and references therein). Therefore, the contribution of leptons must be taken into account. Moreover, not all PeVatrons candidates have been observed in X--rays. In the era of PeVatrons, X-ray observations are mandatory for all PeVatron candidates to support the distinction between hadronic and leptonic emission origins. Given the discovery of TeV haloes in PWNe (leptonic acceleration around pulsars), the crab as a leptonic emitter, and the fact that electrons with energies of $\sim$ 100 TeV can emit gamma rays with similar energies by inverse Compton scattering of Cosmic background photons, the observation of emission up to 1 PeV may not be enough (condition) to claim a PeVatron. Therefore, the true PeVatron definition will be constrained after clarifying the nature of the sub-PeV to PeV emission and through an (intense) discussion in the context of SNR--PWNe topic. HAWC is ready to continually contribute with more powerful detection along with the introduction of outriggers, and the southern observatories such as CTA, H.E.S.S., SWGO, and ALPACA, will provide answers in a richer sky (in SNRs, PWNe, and the Galactic centre).

The link between PeVatrons and astroparticles is crucial because if PeVatrons produce astroparticles, the study of astroparticles can shed light on the nature of the PeVatron itself. The best analysis is to include all three astroparticles in the same study. The relationship between cosmic rays and gamma rays has already been discussed above. The relationship between neutrinos and gamma rays is that the detection of neutrinos is essential for understanding the acceleration of cosmic rays (e.g., Prantic et al.\cite{Prantic2023}). If the detection of neutrinos by gamma rays is confirmed, an accelerator for cosmic rays has been found. Considering the latter and the fact that gamma--ray emission up to 1 PeV is not sufficient to claim a PeVatron, only to propose a PeVatron candidate, a discussion on Pevatrons at ICRC 2021 meeting 55 (ultra-high energy gamma rays and PeVatrons) and a summary between the conveners and the rapporteur, it was discussed that three conditions must be met to claim a Pevatron: 1.- gamma--ray emission indicating acceleration at PeVs, 2.- neutrino coincidence and 3.- molecular gas environment. 

Therefore, without forgetting the meaning of the term PeVatron in the context of cosmic rays, a definition for a PeVatron could be a natural particle accelerator (astronomical object) capable of accelerating particles to PeVs in a molecular environment with neutrino coincidence. In ICRC 2023 and as a relevant topic, the contributions should provide clues to clarify and constrain answers in the new world of PeVatrons.

\section{HAWC meet PeVatrons}

We focus on the Cygnus constellation because three of the 12 LHAASO PeVatrons candidates observed by HAWC are found there: 1.- LHAASO J2108+5157 in the Cygnus OB7 molecular cloud (e.g., delaFuente et al.\cite{delaFuente2023}), 2.- the Cygnus cocoon driven by Cygnus OB2 in the Cygnus X molecular cloud (Abeysekara et al.\cite{Abeysekara2021}), and 3.- the Dragonfly nebula associated with a PWNe towards the Cygnus rift (Abeysekara et al. \cite{Albert2021}). Schneider et al.\cite{Schneider2006} have shown that these molecular clouds and the rift are part of the same region, so the same physical conditions are expected. Moreover, in the first LHAASO catalog, LHAASO J2108+5157 and the Cygnus cocoon are the PeVatrons candidates without SNR or PWNE counterpart association. Besides, the Cygnus cocoon is the only one of the three sources with neutrino detection (e.g., Banik \& Ghosh\cite{Banik2022} and references therein). In Cygnus, the Dragonfly Nebula appears to be a leptonic PeVatron, and LHAASO J2108+5157 turns out to be a deeply enigmatic PeVatron, being the only one in a vast molecular region where there is no clear counterpart, not even in the star--forming region Kronberger 82 (delaFuente et al.\cite{delaFuente2023}). 

Although several studies search for a leptonic source and the TeV and sub(PeV) emissions suggest a leptonic nature, Toledano--Ju\'arez\cite{TJ2023} with optically thin $^{13}$CO(1$\rightarrow$0) arcseconds observations taken with Nobeyama's 45 m radio-telescope (resolution of 26 arcsecond), show that a molecular cloud matches in morphology the extended Fermi--LAT emission (above 2 GeV), suggesting that this molecular cloud is the target of gamma rays from an unknown PeVatron. They suggest that this possible PeVatron, if nearby, could be embedded in this molecular gas. To avoid controversy and polemics, we can say that HAWC meets PeVatrons in the Cygnus cocoon because the three conditions mentioned above for a PeVatron are most likely fulfilled there.

\section*{Acknowledgments}
The following link \url{https://www.hawc-observatory.org/collaboration/} contains the HAWC general acknowledgements. The HAWC collaboration also thanks  CONACyT grant: CF-2023-I-645. EdelaF thanks the Inter-University Research Programme of the Institute for Cosmic Ray Research, University of Tokyo, grant 2023i--F--005, and doctorado de Ciencias F\'isico--Matem\'aticas of the Centro Universitario de los Valles, Universidad de Guadalajara for financial support. He also thanks colegio departamental de F\'isica, Centro Universitario de Ciencias Exactas e Ingener\'ias, CUCEI, for support in several research stays, and the administrative and academic staff of the Institute of Cosmic Ray of the University of Tokyo, for several supports during his Sabbatical in 2021.  IT--J acknowledges support from Consejo Nacional de Ciencias y Tecnolog\'ia, M\'exico (CONACyT) grant 754851. We thank Kato Sei of the Institute of Cosmic Ray of the University of Tokyo for his peer-review including constructive comments and suggestions that helped to improve the manuscript.

\clearpage
\section*{Authors List of the HAWC Collaboration}
\small
\noindent
A.U. Abeysekara$^{48}$,
A. Albert$^{21}$,
R. Alfaro$^{14}$,
C. Alvarez$^{41}$,
J.D. \'Alvarez$^{40}$,
J.R. Angeles Camacho$^{14}$,
J.C. Arteaga-Vel\'azquez$^{40}$,
K. P. Arunbabu$^{17}$,
D. Avila Rojas$^{14}$,
H.A. Ayala Solares$^{28}$
% ...
\\

\scriptsize
% ...
$^{14}$Instituto de F\'isica, Universidad Nacional Aut\'onoma de M\'exico, Ciudad de M\'exico, M\'exico,
% ...
$^{17}$Instituto de Geof\'isica, Universidad Nacional Aut\'onoma de M\'exico, Ciudad de M\'exico, M\'exico,
% ...
$^{21}$Physics Division, Los Alamos National Laboratory, Los Alamos, NM, USA,
% ...
$^{28}$Department of Physics, Pennsylvania State University, University Park, PA, USA,
% ...
$^{40}$Instituto de F\'isica y Matem\'aticas, Universidad Michoacana de San Nicol\'as de Hidalgo, Morelia, Michoac\'an, M\'exico,
$^{41}$FCFM-MCTP, Universidad Aut\'onoma de Chiapas, Tuxtla Guti\'errez, Chiapas, M\'exico,
% ...
$^{48}$Department of Physics and Astronomy, University of Utah, Salt Lake City, UT, USA,
% ...

\end{document}